\documentclass[hyper, notoc]{JHEP3} 
\usepackage[dvips]{graphicx}
\usepackage{youngtab}   
  
\newcommand{\be}{\begin{equation}}  
\newcommand{\ee}{\end{equation}}  
\newcommand{\bea}{\begin{eqnarray}}     
\newcommand{\eea}{\end{eqnarray}}     
\newcommand{\ov}{\overline}     
\newcommand{\lra}{\leftrightarrow}

\title{A Fat Higgs with a Fat Top}   
\author{Antonio Delgado\\
TH-Division, CERN, 1211 Geneva 23, Switzerland
\\Department of Physics and 
Astronomy, Johns  Hopkins University, Baltimore, MD 21218}  
\author{Tim M.P. Tait \\ Fermi National Accelerator Laboratory, 
Batavia, IL 60510 \\
HEP Division, Argonne National Lab, Argonne IL 60439 }  
 
\abstract{A new variant of the supersymmetric Fat Higgs model is presented
in which the MSSM Higgses as well as the top quark are composite.  The
underlying theory is an $s$-confining $SU(3)$ gauge theory with the MSSM
gauge groups realized as gauged sub-groups of the chiral flavor symmetries.
This motivates the large Yukawas necessary for the large top mass and
SM-like Higgs of mass $\gg M_Z$ in a natural way as the residual of the
strong dynamics responsible for the composites.  This removes fine-tuning
associated with these couplings present in the original Fat Higgs and
``New Fat Higgs'' models, respectively.} 
 
\keywords{aft, sub, suy}  
\preprint{ANL-HEP-PR-05-18\\
CERN-TH/2005-056\\
FERMILAB-PUB-05-028-T\\}
  
\begin{document}  
  
\section{Introduction}  
\label{sec:intro} 

Supersymmetry (SUSY) is the most cherished and best studied vision of physics
beyond the Standard Model (SM).  SUSY tames the quadratic divergences that
destabilize the electroweak (EW) scale, and results in a host of new particles
which should be discovered in the near future if the SUSY vision of particle
physics should prove correct.

However, LEP-II has left the minimal supersymmetric standard model (MSSM)
in an interesting situation \cite{Abbaneo:2001ix}.  
The minimal model predicts a light Higgs
whose tree-level mass is at most $M_Z$, in contradiction with the LEP-II
limit of $M_h^{(SM)} \geq 115$ GeV.
In order to survive the LEP limit, one must either invoke very large radiative
corrections from the top sector \cite{Carena:1995wu}, 
CP violation chosen in a very particular way \cite{Carena:2002bb}, or
abandon the minimal model in favor of more ingredients 
\cite{Ellis:1988er,Batra:2003nj,Batra:2004vc,Harnik:2003rs,Chang:2004db,Casas:2003jx,Maloney:2004rc}.
The invocation of large radiative corrections is particularly troublesome,
because this tends to introduce unacceptably large corrections to the EW
scale, recreating a ``little hierarchy problem''.  
While there is some uncertainty in the estimates for the lightest CP even
Higgs mass originating in
the uncertainty in measured top mass, it appears that the MSSM requires
fine-tuning at the level of a few per cent if it is to be consistent with
LEP data, and is uncomfortably fine-tuned.  This is the ``Supersymmetric
Little Hierarchy Problem''.

The Fat Higgs (FH) \cite{Harnik:2003rs} is a particular, 
interesting solution to this dilemma.  It proposes an alternative to
the standard MSSM picture of electroweak symmetry breaking (EWSB) and results
in a heavier ``light'' CP-even Higgs than can be realized in that standard
scenario, thus naturally evading the LEP-II bounds.  It originates from an
$s$-confining theory, 
in which a number of fundamental preons charged
under a strong $SU(2)$ form Higgs bosons as composites.  A variation
\cite{Chang:2004db} has a composite singlet from
an $s$-confining $SU(4)$ theory, but the EWSB Higgses are 
fundamentals.  Both theories have interesting distinctive SUSY Higgs
phenomenology \cite{Batra:2004vc,Harnik:2003rs}, 
largely due to the fact that the Higgs quartic interaction
may be much larger than is suggested by perturbative unification
\cite{Haber:1986gz}.

Both of these FH theories are challenged in producing large Yukawa 
interactions.  The original FH must generate fermion masses through Yukawa
interactions which couple the composite $H$ and $\overline{H}$
to the fundamental quarks and
leptons.  At the level of the preons, this is a 
non-renormalizable super-potential
coupling, which the original FH generates 
from renormalizable interactions by integrating out a pair of
Higgs-like fields uncharged under the strong $SU(2)$ 
(see Figure~\ref{fig:yukawa}).  
The resulting Yukawas thus depend on fundamental parameters as,
\bea
y_{eff} & \sim & \frac{y y^\prime}{4 \pi} \frac{\Lambda}{M_{H}}
\eea
in which $y$, $y^\prime$ are Yukawas between the preons and/or fundamental
fermion superfields (at the compositeness scale $\Lambda$), $4 \pi$ 
is the naive dimensional analysis (NDA) counting
\cite{Cohen:1997rt} for the coupling of a composite to fundamental fields,
 $\Lambda$ is the scale of $s$-confinement 
of the strong $SU(2)$ and $M_{H}$ 
is the (supersymmetric) mass of the Higgs-like fields.
For the light fermions, this is not problematic.  Small fermion masses are
easily realized.  For the top quark, producing a Yukawa coupling of order one
requires tuning the scales $\Lambda$ and $M_{H}$ to be close to one another
(which is somewhat counter-intuitive since they are in principal unrelated to
one another, though it was argued in
\cite{Harnik:2003rs} that the coincidence of scales could arise from a
flavor symmetry) and that the underlying $y$ and $y^\prime$ 
be large at $\Lambda$ to compensate for
the $4 \pi$.  This last fact is also potentially
a source of fine-tuning.  The strong
$SU(2)$ tries to renormalize $y$ and $y^\prime$ strong at low energies.
This is helpful in that it compensates the suppression, but dangerous
because a large super-potential coupling may ruin the conformal regime
of the theory above $\Lambda$.

While it is possible that interesting (and phenomenologically viable)
low energy dynamics would emerge
in this case, the additional strong $y$ (and/or in generalizations
$y^\prime$) couplings potentially disrupt the low energy 
$s$-confinement solution, and makes it difficult to draw firm conclusions
about the low energy physics.  One is thus forced to assume that $y$
and $y^\prime$ become moderately strong, but do not quite reach truly strong
coupling before the $s$-confinement scale.  Another way
to consider the tension is to note\footnote{We are indebted to
Kaustubh Agashe for discussions on this point.} that one must
tune the original $y$ and $y^\prime$ to some very particular values in the
UV such that they become large enough (but not too large) at $\Lambda$.
The ``New Fat Higgs'' \cite{Chang:2004db} avoids this issue for the top
Yukawa, because in that case the EW Higgses and the quarks are fundamental.
Thus, the strong $SU(4)$ does not effectively drive that interaction strong
at low energies.  However, it recreates the problem for the Higgs quartic
itself, because now the quartic links the composite EW singlet $S$ to
the fundamental EW Higgses $H$ and $\ov{H}$, and thus
feels the same sort of tension when one tries to obtain a large Higgs
quartic.

\FIGURE[t]{\includegraphics[width=3.5in]{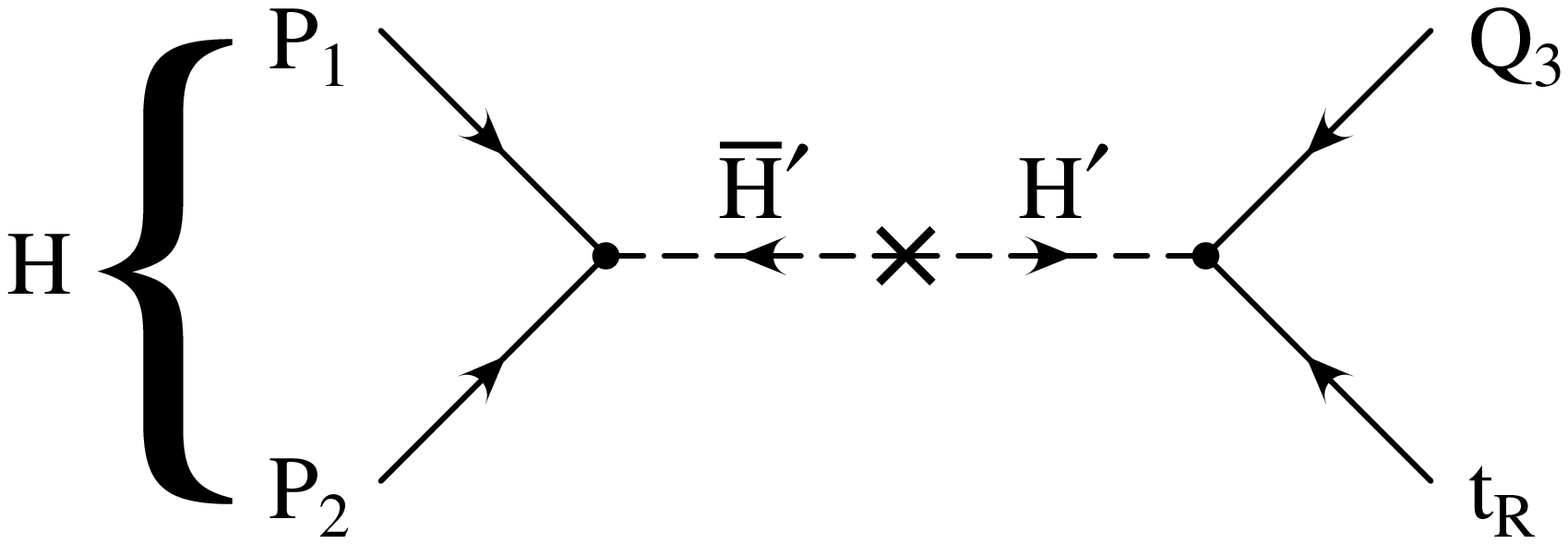} 
\label{fig:yukawa}\caption{Example graph for how the top Yukawa coupling
is generated in the Fat Higgs model by integrating out a pair
of Higgs-like superfields ($\ov{H}^\prime$, $H^\prime$) to generate
a non-renormalizable interaction between preons ($P_1$ and $P_2$)
bound into a composite Higgs $H$. }}

In this article, we explore a new incarnation of the Fat Higgs.  Our theory
is an $SU(3)_s$ SUSY gauge theory which $s$-confines, producing a composite
singlet $S$ and doublets $H$ and $\ov{H}$ as in the original Fat Higgs.
However, the additional
preons are arranged such that they also produce a composite
third generation quark doublet ($Q_3$) and up-type singlet ($t_R$).  The
dynamically generated super-potential contains the terms needed for FH-style
EWSB, but it also includes the top Yukawa coupling\footnote{For
pre-Fat Higgs SUSY models which realize the large top Yukawa coupling
through $s$-confining dynamics, see \cite{Strassler:1995ia}.}.  
Since all fields
requiring large Yukawa interactions are composite, we have removed
the need for strong underlying Yukawa interactions, and thus the danger that
the low energy physics could be spoiled by out-of-control non-perturbative
couplings.  Furthermore, while we will still
need to invoke massive fields to generate
the Yukawa interactions of the light fermions, there is considerably less need
to fine-tune the mass of these ``spectator'' superfields 
($M_{H}$) to the
$s$-confinement scale $\Lambda$, and/or invoke underlying super-potential
couplings which are dangerously large.
  
In Sec.~\ref{sec:model}, we present the model and show how it gives rise to
all of the required low energy structure of the MSSM.  In Sec.~\ref{sec:uni},
we address some of the issues regarding high energy gauge coupling unification.
In Sec.~\ref{sec:pheno} we discuss some of the distinctive phenomenology.
And in Sec.~\ref{sec:concl} we conclude.

\section{An $SU(3)$ Model} 
\label{sec:model} 

\begin{table}
\centering
\begin{tabular}{lccccc}
                          & $SU(3)_s$            & $SU(3)_c$            & $SU(2)_W$    & $U(1)_Y$  & $Z_2$ 
\\ \hline
$P_3$                     & \yng(1)              & \yng(1)              & $\mathbf{1}$ & $0$       & $+$ \\
$P_1$                     & \yng(1)              & $\mathbf{1}$         & $\mathbf{1}$ & $-2/3$    & $-$ \\
$\ov{P}_2$          & $\ov{\yng(1)}$ & $\mathbf{1}$         & \yng(1)      & $+1/6$    & $-$ \\
$\ov{P}_1$          & $\ov{\yng(1)}$ & $\mathbf{1}$         & $\mathbf{1}$ & $+2/3$      & $+$ \\
$\ov{P}_{\tilde 1}$ & $\ov{\yng(1)}$ & $\mathbf{1}$         & $\mathbf{1}$ & $-1/3$ & $-$ \\
\hline
$P^\prime$                & \yng(1)              & $\mathbf{1}$         & $\mathbf{1}$ & $+1/3$    & $-$ \\
$\ov{P}^\prime$     & $\ov{\yng(1)}$ & $\mathbf{1}$         & $\mathbf{1}$ & $-1/3$    & $-$ \\ 
\end{tabular}
\caption{The $SU(3)_s$-charged Preons.   The first set are 
those participating in the $s$-confining phase.  The second category are 
integrated out, triggering $s$-confinement.}
\label{tab:preons}
\end{table}

Our model has an extended gauge symmetry,
\bea 
SU(3)_s \times SU(3)_c \times SU(2)_W \times U(1)_Y .  
\eea
$SU(3)_s$ is a 
``strong'' group which will be responsible for generating the MSSM Higgses,
a Fat-Higgs like singlet, and top from a set of preons, and the 
remaining gauge groups are as in the
MSSM.  The particle content charged under $SU(3)_s$ consists of a set of 
preons listed in Table~\ref{tab:preons}.  Since the matter is vector-like
with respect to $SU(3)_s$, we follow the usual fashion and
refer to it as a ``SUSY QCD'' theory, but this
should not be confused with the usual color interaction of the MSSM,
$SU(3)_c$.
Note that the MSSM gauge groups
are gauged sub-groups of the $SU(F) \times SU(F) \times U(1)_B$ chiral
symmetries.
The set of preons is non-anomalous (in fact, it is vector-like)
with respect to $SU(3)_s$, and there are no mixed anomalies between
$SU(3)_s$ and the MSSM gauge groups.  However, the MSSM gauge symmetries
are anomalous with respect to themselves.  This is in 
fact related to the point that the strong sector will eventually give rise to
a composite $Q_3$, $t_R$, $\ov{H}$, $S$ and $H$, but not to 
$b_R$, $L_3$, or $e_3$.  
Thus, we introduce a set of fundamental fields 
uncharged under $SU(3)_s$ in Table~\ref{tab:fundamentals}.  The first 
and second generation superfields appear as fundamental fields, 
as in the MSSM.  Also indicated are the charges of the fields under
a $Z_2$ ``$R$-parity'' which plays the same role to suppress dangerous
renormalizable
baryon- and lepton-number violating processes as it does in the MSSM.
The assignment of preon hypercharges is not completely determined by
requiring the correct hypercharges for the composites, and 
the particular choice we make is based partly on aesthetics 
(requiring that all exotic colored particles have 
charges $\pm 1/3$ or $\pm 2/3$
and all exotic uncolored particles have charges $\pm 1$ or zero),
and partly motivated by gauge coupling unification as we shall see below.
Many fundamental Yukawa interactions can be formed out
of these fields. To preserve readability, we discuss these 
in groups in the subsections below.

\begin{table}[[t]
\centering
\begin{tabular}{lccccc}
                      & $SU(3)_s$    & $SU(3)_c$              & $SU(2)_W$        & $U(1)_Y$  & $Z_2$ 
\\ \hline
$L_i$                 & $\mathbf{1}$ & $\mathbf{1}$           & {\yng(1)}        & $-1/2$    & $-$ \\
$e_i$                 & $\mathbf{1}$ & $\mathbf{1}$           & $\mathbf{1}$     & $+1$      & $-$ \\
$Q_{1,2}$             & $\mathbf{1}$ & $\yng(1)$              & $\yng(1)$        & $+1/6$    & $-$ \\
$d_i$                 & $\mathbf{1}$ & {$\ov{\yng(1)}$} & $\mathbf{1}$     & $+1/3$    & $-$ \\
$u_{1,2}$             & $\mathbf{1}$ & {$\ov{\yng(1)}$} & $\mathbf{1}$     & $-2/3$    & $-$ \\
$\ov{q}_1$      & $\mathbf{1}$ & {$\ov{\yng(1)}$} & $\mathbf{1}$     & $-2/3$      & $+$ \\
$\ov{q}_2$      & $\mathbf{1}$ & {$\ov{\yng(1)}$} & $\mathbf{1}$     & $+1/3$ & $-$ \\
\hline
$H^\prime$            & $\mathbf{1}$ & $\mathbf{1}$           & \yng(1)          & $+1/2$    & $+$ \\
$\ov{H}^\prime$ & $\mathbf{1}$ & $\mathbf{1}$           & \yng(1)          & $-1/2$    & $+$ \\
\end{tabular}
\caption{Additional fundamental fields 
for the $SU(3)$ model.  The index $i=1,2,3$ denotes the usual 
generation number.}
\label{tab:fundamentals}
\end{table}

This theory is SUSY $SU(3)$ QCD with $5$ flavors, which is inside the 
conformal window \cite{Terning:2003th}.  
From any value of the $SU(3)_s$ gauge coupling
at very high scales, it flows (assuming, as we will do so, that all of the
fundamental Yukawa interactions are not strong enough to disrupt the
approximate scale-invariance) at lower scales to the fixed point at,
\bea
g_*^2 & \simeq & \frac{4 \pi^2}{3}
\eea
We include a super-potential mass for $P^\prime$ 
(and for the uncolored $H^\prime$),
\bea
W_m = M_P \ov{P}^\prime P^\prime + M_H \ov{H}^\prime H^\prime
.
\eea
Below $M_P$, the $P^\prime$, $\ov{P}^\prime$
flavor may be integrated out and the theory loses
conformality, flowing to an $s$-confining phase \cite{Csaki:1996zb}.  
We denote
the confinement scale by $\Lambda$, and estimate from the large fixed point
coupling $g_*$ that the two scales are approximately equal,
\bea
\Lambda & \simeq & M_P ~.
\eea
The scale $M_P$ must be input by hand, and 
determines the strong coupling scale $\Lambda$.

\TABLE[t]{
\begin{tabular}{lccccc}
                                      &  & $SU(3)_c$            & $SU(2)_W$        & $U(1)_Y$  & $Z_2$ 
\\ \hline
$B_1 \lra t_R$ &  $P_3 P_3 P_1$         & $\ov{\yng(1)}$ & $\mathbf{1}$     & $-2/3$    & $-$ \\
$B_2 \lra S$ & $ P_3 P_3 P_3$           & $\mathbf{1}$         & $\mathbf{1}$     & $0$       & $+$ \\
$\ov{B}_1 \lra H$ & $ \ov{P}_2 \ov{P}_1 \ov{P}_{\widetilde{1}}$     & $\mathbf{1}$         & \yng(1)          & $+1/2$    & $+$ \\
$\ov{B}_2 \lra \psi$ & $\ov{P}_2 \ov{P}_2 \ov{P}_1$     & $\mathbf{1}$         & $\mathbf{1}$     & $+1$ & $+$ \\
$\ov{B}_3 \lra \chi$ & $\ov{P}_2 \ov{P}_2 \ov{P}_{\widetilde{1}}$     & $\mathbf{1}$         & $\mathbf{1}$     & $0$ & $-$ \\
$M_1 \lra Q_3$ & $P_3 \ov{P}_2$            & \yng(1)              & \yng(1)          & $+1/6$    & $-$  \\
$M_2 \lra {q}_1$ & $P_3 \ov{P}_1$           & \yng(1)              & $\mathbf{1}$     & $+2/3$       & $+$  \\
$M_3 \lra {q}_2$  & $P_3 \ov{P}_{\widetilde{1}}$          & \yng(1)              & $\mathbf{1}$     & $-1/3$ & $-$  \\
$M_4 \lra \ov{H}$ & $P_1 \ov{P}_2$    & $\mathbf{1}$         & $\yng(1)$        & $-1/2$    & $+$  \\
$M_5 \lra \ov{\chi}$ & $P_1 \ov{P}_1$    & $\mathbf{1}$         & $\mathbf{1}$     & $0$  & $-$  \\
$M_6 \lra \ov{\psi}$ & $P_1 \ov{P}_{\widetilde{1}}$    & $\mathbf{1}$         & $\mathbf{1}$     & $-1$  & $+$
\end{tabular}
\hspace*{5in}
\label{tab:composites}
\caption{Composites of the $SU(3)$ model.}}

\subsection{Composites and Dynamical Super-potential}

Below the confinement scale, the theory can be described by composite
$SU(3)_s$-invariant mesons ($M$) and baryons ($B$, $\ov{B}$), listed in 
Table~\ref{tab:composites}.  A dynamical super-potential is generated
with form,
\bea
W_{dyn} & = &
\frac{1}{\Lambda^5} \left\{ \ov{B} M B - {\rm det}~M \right\} 
\nonumber \\
& \rightarrow & \lambda \left\{ H Q_3 t_R + H \ov{H} S 
+ \psi {q}_2 t_R + \psi \ov{\psi} S + \chi \ov{\chi} S
+ \chi {q}_1 t_R
- \frac{\lambda}{\Lambda} {\rm det} M \right\} ,
\label{eq:wdyn}
\eea
where in the second line
we rescaled the baryons and mesons to canonically normalized 
superfields.  
It will not be very important for our purposes, but we note for completeness
that one may express the irrelevant interactions as,
\bea
{\rm det}~M & = & \epsilon_{ij} \epsilon_{\alpha \beta \gamma}
\left(
\ov{H}^i Q_3^{\alpha j} q_1^\beta q_2^\gamma +
\ov{\chi} Q_3^{\alpha i} Q_3^{\beta j} q_2^\gamma + 
\ov{\psi} Q_3^{\alpha i} Q_3^{\beta j} q_1^\gamma 
\right) ,
\eea
suppressed by the confinement scale $\lambda / \Lambda$.
We have also
provided the naive dimensional analysis (NDA) estimate for the coupling
$\lambda \sim 4\pi$ \cite{Cohen:1997rt}.  
Thus, this model dynamically generates the Fat Higgs sector and
super-potential, along with the
top Yukawa coupling and some exotic interactions with exotic
superfields.  Note that the exotics occur in pairs in these interactions,
because they arise exclusively from composites which include an odd
number of $\ov{P}_1$ and $\ov{P}_{\widetilde{1}}$.

We shall see below that $q_1$ and $q_2$ 
receive masses of order $\Lambda$.  Thus, below $\Lambda$ the
relevant couplings in (\ref{eq:wdyn}) are the top Yukawa $y_t$, the 
$S H \ov{H}$ interaction $\lambda_H$, the $S \psi \ov{\psi}$ 
interaction $\lambda_\psi$,
and the $S \chi \ov{\chi}$ interaction $\lambda_\chi$.  
All of these are equal and
of order $\lambda \sim 4 \pi$ 
at the scale $\Lambda$, but because the $q's$ decouple
at that scale, and because of our having gauged subgroups of the chiral 
symmetries of the SUSY QCD theory, they evolve apart at lower energies.

\FIGURE[t]{\includegraphics[width=5.0in]{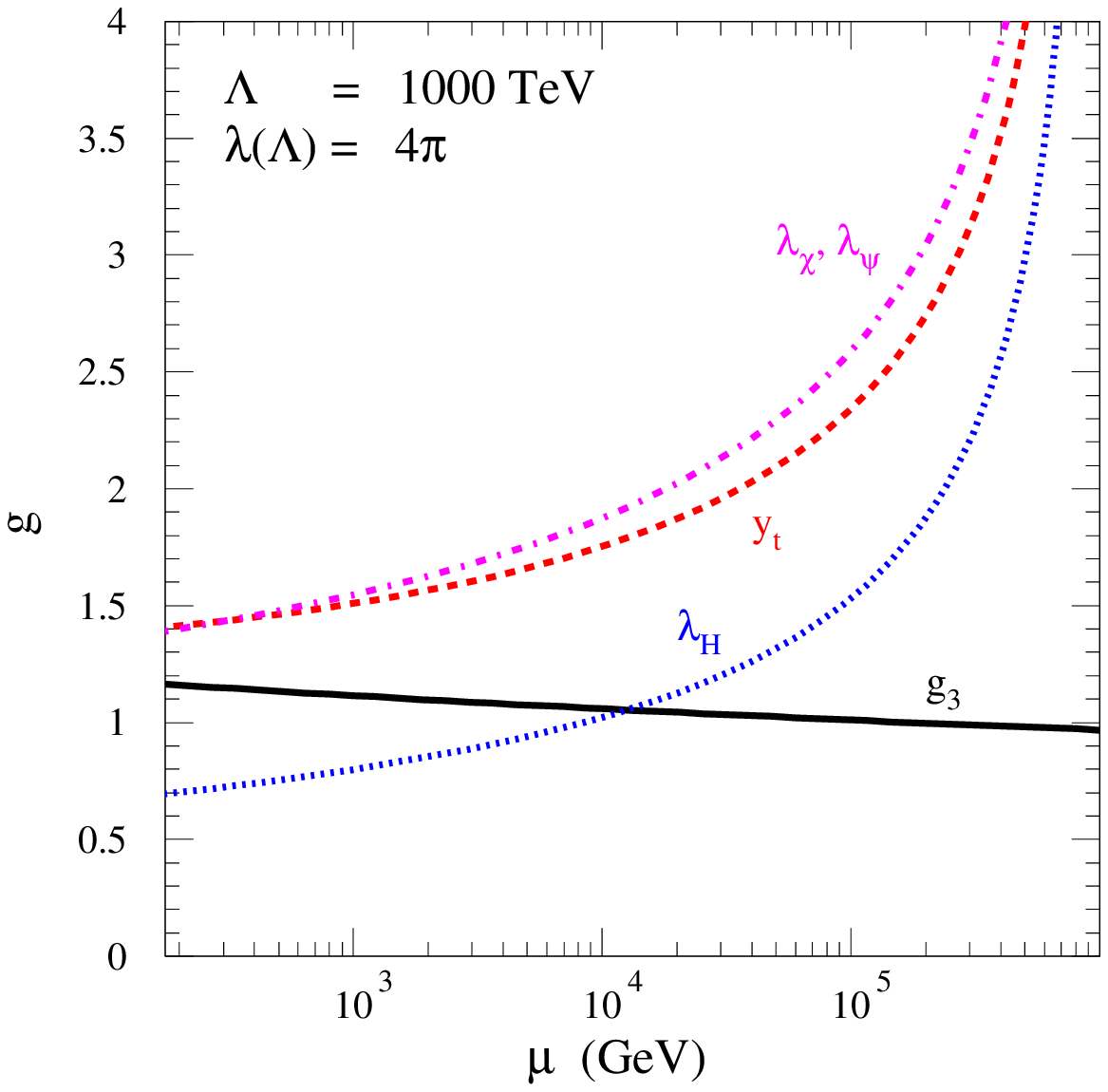} 
\label{fig:rge}\caption{The RGE evolution from $\lambda = 1000$ TeV to
$v$ of the strong coupling $g_3$ (solid curve), top Yukawa interaction
$y_t$ (dashed curve), $S H \ov{H}$ interaction $\lambda_H$ (dotted curve),
and $S \psi \ov{\psi}$ and $S \chi \ov{\chi}$ interactions
$\lambda_\psi$ and $\lambda_{\chi}$ (dot-dashed curve). }}

In order to discuss the top mass and EWSB, these should be evolved down to
energy scales of order the electroweak scale $v$.  
At one-loop, below $\Lambda$, the dominant renormalization effects are
from $y_t$, and $\lambda_{(H,\psi,\chi)}$ 
themselves, and from the $SU(3)_c$ coupling
$g_3$.  The one loop renormalization group equations (RGEs) are
\bea
\frac{dg_3}{dt} & = & -\frac{3}{16\pi^2} g_3^3 \\
\frac{dy_t}{dt} & = &  \frac{y_t}{16\pi^2} 
\left[ 6 |y_t|^2 + |\lambda_H|^2 - \frac{16}{3} g_3^3 \right] \\
\frac{d\lambda_H}{dt} & = &  \frac{\lambda_H}{16\pi^2} 
\left[ 3 |y_t|^2 + 4 |\lambda_H|^2 + |\lambda_\psi|^2 
+ |\lambda_\chi|^2\right] \\
\frac{d\lambda_\psi}{dt} & = &  \frac{\lambda_\psi}{16\pi^2} 
\left[ 2 |\lambda_H|^2 + 3 |\lambda_\psi|^2 + |\lambda_\chi|^2 \right] \\
\frac{d\lambda_\chi}{dt} & = &  \frac{\lambda_\chi}{16\pi^2} 
\left[ 2 |\lambda_H|^2 + 3 |\lambda_\chi|^2 + |\lambda_\psi|^2 \right]
\eea
where $t$ is the renormalization scale $t \equiv \log \mu_R$.  
Since $\lambda_\psi = \lambda_\chi$ at scale $\Lambda$, 
these coupling strengths
will remain equal up to very small effects from the different hypercharges
of $\psi$ and $\chi$.

The fact that the top mass  has been measured at the 
Tevatron \cite{Hill:2004qu} allows us
to approximately fix $\Lambda$, up to the choice of $\tan \beta$.  As
values of $\tan \beta \sim 1$ result in the largest light CP even Higgs
masses, we make this choice for which the target $y_t$ is about $\sqrt{2}$.
Solving the coupled equations numerically and imposing
this requirement fixes $\Lambda \sim 10^4 \times v$ 
(i.e. $\Lambda \sim 1000$~TeV), and predicts that
$\lambda_H$ will be somewhat less than $y_t$ itself.  
An example is shown in figure~\ref{fig:rge}.
Note that there are order one
uncertainties in $\lambda(\Lambda)$, which could easily
modify our estimate for $\Lambda$ by an order of 
magnitude\footnote{There are also order one uncertainties in the RGE
evolution from higher orders close to scale $\Lambda$, where
the couplings are strong, as well.}.  
Irrespectively, the 
prediction that the Higgs quartic is approximately locked to the top Yukawa
interaction is an interesting feature of the model.


\subsection{Electroweak Symmetry Breaking}

We include a Yukawa coupling in the fundamental theory,
\bea
W_S & = & -y_S \epsilon_{\alpha \beta \gamma} 
P_3^{\alpha} P_3^{\beta} P_3^{\gamma} 
\nonumber \\
& \rightarrow & -\left(\frac{y_S}{4 \pi} \Lambda^2 \right) S ~,
\eea
(where $\alpha$, $\beta$, and $\gamma$ are $SU(3)_c$ indices, and
the $SU(3)_s$ indices are similarly contracted anti-symmetrically
but not shown for clarity)
which becomes a tadpole for $S$ below $\Lambda$.
Combined with $W_{dyn}$, this results in Higgs super-potential,
\bea
W_H & = & \lambda_H S \left( H \ov{H} - v_0^2 \right)
+ \lambda_\psi S \psi \ov{\psi} + \lambda_\chi S \chi \ov{\chi} 
\eea
where $v_0^2$ has NDA estimate (at scale $\Lambda$),
\bea
v_0^2 & \sim & \frac{y_S}{\lambda \left(4 \pi\right)} \Lambda^2
\sim
\frac{y_S}{\left(4 \pi\right)^2} \Lambda^2
\eea
thus indicating that $v_0$ is naturally at least an order of 
magnitude below $\Lambda$, and will be smaller if $y_S$ takes
a sufficiently small value (as we will assume it does in order to
appropriately generate the EW scale).  
Aside from the presence of the additional
superfields $\psi$, $\ov{\psi}$, $\chi$, $\ov{\chi}$,
this is the super-potential of the Fat Higgs, leading
to a electroweak symmetry-breaking even in the supersymmetric limit.

The scalar Higgs
potential consists of the contribution from the dynamical super-potential
above, the MSSM $D$-terms, and the corrections from soft SUSY breaking.
There is also an effective  $\mu$ term induced by integrating out
$H^\prime$ and $\ov{H}^\prime$ as described below in 
section~\ref{sec:residual}.
Altogether, this leads to a scalar potential,
\bea
V_H & = &
| \lambda_H H \ov H + \lambda_\psi \psi \ov{\psi} 
+ \lambda_\chi \chi \ov{\chi} - v_0^2 |^2 
+ \lambda_H^2 \left( | S H |^2 + |S \ov H|^2 \right)
\nonumber \\ & &
+ \lambda_\psi^2 \left( | S \psi |^2 + |S \ov \psi|^2 \right)
+ \lambda_\chi^2 \left( | S \chi |^2 + |S \ov \chi|^2 \right)
\nonumber \\ & &
+ \frac{g_2^2}{8} \left( H^\dagger \vec{\tau} H + 
\ov{H}^\dagger \vec{\tau} \ov{H} \right)^2
+ \frac{g_1^2}{2} \left( \frac{1}{2} |H|^2 
- \frac{1}{2} |\ov{H}|^2 
+ |\psi|^2 - |\ov{\psi}|^2
\right)^2
\nonumber \\ & &
+ \left( m_H^2 + |\mu|^2 \right) |H|^2 
+ \left( m_{\ov{H}}^2 + |\mu|^2 \right) |\ov{H}|^2 + m_S^2 |S|^2
\nonumber \\ & &
+ m_\psi^2 |\psi|^2 + m_{\ov{\psi}}^2 |\ov{\psi}|^2
+ m_\chi^2 |\chi|^2 + m_{\ov{\chi}}^2 |\ov{\chi}|^2
\nonumber \\ & &
+ \left\{ A_{S} \left( \lambda_H S H \ov{H} + \lambda_\psi S \psi \ov{\psi}
+ \lambda_\chi S \chi \ov{\chi} \right) - T_S v_0^2 S + h.c. \right\} ~~,
\eea
where $g_{1,2}$ are the MSSM $U(1)/SU(2)$ gauge couplings, and the
$m$'s, $A_S$, and $T_S$ are soft SUSY breaking
parameters.  We have assumed that the $A$ terms are locked together by the
underlying chiral symmetries of the SUSY QCD theory, and in the same spirit
ignored other potential SUSY breaking terms such as 
$B \mu$-like terms involving $H \ov{H}$,
$\psi \ov{\psi}$, and $\chi \ov{\chi}$.
Of course, we expect that the equality of the $A$ terms is only approximate,
as the RGEs will split them apart just as it does the $\lambda$ interactions,
but we continue to neglect such splittings to simplify the discussion.

In general, the minimization conditions are 
quite complicated, but we sketch a solution below.  To simplify matters,
we begin by considering $m_H = m_{\ov H} = m_S \equiv m$,
$m_\psi = m_{\ov{\psi}} = m_\chi = m_{\ov{\chi}} \equiv M$, 
$A_H = A_\psi = A_\chi = T_S = 0$, and ignore the MSSM $D$-terms.  
We will consider
deviations from these assumptions below.  Under these conditions, the potential
is symmetric under $H \lra \ov{H}$ and 
$\psi \lra \ov{\psi} \lra \chi \lra \ov{\chi}$.  The SM-like Higgs is
$h=(H^0 + \ov{H}^0) / \sqrt{2}$, and we denote the common 
vacuum expectation value (VEV) of
$\psi$, $\ov{\psi}$, $\chi$, and $\ov{\chi}$ as $\phi/\sqrt{2}$.  
The scalar potential becomes
\bea
\left( 
\frac{\lambda_H^2}{4}h^4 + \lambda_\psi \phi^4 
+ \lambda_H \lambda_\psi h^2 \phi^2 + 2 \lambda_\psi^2 |S|^2 \phi^2 \right)
+ m^2 |S|^2 
\nonumber \\
+ \left( m^2 + |\mu|^2 - \lambda_H^2 v_0^2 \right) h^2
+ \left( M^2 - \lambda_\psi^2 v_0^2 \right) \phi^2 
\eea
and the vacuum crucially depends on the signs of the quantities
$(m^2 + |\mu|^2 - \lambda_H^2 v_0^2)$ 
and $(M^2 - \lambda_\psi^2 v_0^2)$.  Under the 
relatively mild requirement that the soft masses respect,
\bea
\left( m^2 +|\mu|^2 - \lambda_H^2 v_0^2 \right) & < & 0 \\
\label{eq:ewsbreq}
\left( M^2 - \lambda_\psi^2 v_0^2 \right) & > & 0
\eea
we arrive at the solution 
$\langle H \rangle = \langle \ov{H} \rangle = \sqrt{v_0^2 - ( m^2 + |\mu|^2 )/ \lambda_H^2}$, $\langle S \rangle = \langle \psi \rangle =  \langle \ov{\psi} \rangle =  \langle \chi \rangle = \langle \ov{\chi} \rangle = 0$, leading to
viable\footnote{Note that a VEV for $\psi$ or $\ov{\psi}$
would lead to large (tree level) corrections to $\Delta \rho$.} EWSB.
Including the $D$ terms and relaxing the universality among the soft masses
will not disrupt this general feature, provided
$m_\psi$, $m_{\ov{\psi}}$, $m_\chi$, and $m_{\ov{\chi}}$ continue to
individually 
satisfy Eq.~(\ref{eq:ewsbreq}), though it will modify the expressions
for the VEVs and cause 
$\tan \beta \equiv \langle H \rangle / \langle \ov{H} \rangle$ 
to deviate from unity.

We also consider non-zero values for $A_S$ and $T_S$.  Both of these
terms, combined with the EWSB VEVs for $H$ and $\ov{H}$, generate
tadpoles for $S$ which will generically result in it acquiring a VEV of order
the weak scale, and further complicating the precise relation between the
underlying parameters and $\langle H \rangle$ and $\langle \ov{H} \rangle$.
The VEV for $S$ is crucial, because combined with the dynamical 
super-potential, it provides supersymmetric masses for the 
fermionic components\footnote{Alternately, one may introduce
further spectators to marry $\psi$, $\ov{\psi}$, $\chi$, and $\ov{\chi}$
with masses of order $\Lambda$ through non-renormalizable operators mediated
by a new set of spectator preons.  While this results in a more minimal
particle content below $\Lambda$ (and reproduces precisely the FH scalar
potential), it requires many more ingredients, and
thus we prefer to accept the extra light states at the weak scale.}
of $\psi$, $\ov{\psi}$, $\chi$ and $\ov{\chi}$.  Thus,
we expect that in generic points in the parameter space, subject to quite
mild constraints, phenomenologically
viable EWSB and weak scale masses for the uncolored exotics result.

\subsection{Light Fermion masses} 

We have seen that the top Yukawa coupling and Higgs quartic are 
generated by the strong dynamics, and are naturally large. 
The remainder of the fermion masses can also be generated in
the following ways.

\subsubsection{Charged Leptons}

The lepton sector is entirely fundamental, so the required operators are
dimension 5 at the preon level, to connect $L_i$, $e_j$ and the composite
Higgs $\ov{H}$.  The needed underlying interactions are generated
by integrating out the spectators $H^\prime$ 
and $\ov{H}^\prime$ (just as in the original FH), and result in,
\bea
W_L & = & y_{H^\prime} H^\prime P_1 \ov{P}_2
+ y^e_{ij} \ov{H}^\prime L_i e_j
\nonumber \\
& \rightarrow &
\left( \frac{y^e_{ij} y_{H^\prime}}{4 \pi} \frac{\Lambda}{M_H} \right)
\ov{H} L_i e_j ~.
\label{eq:ye}
\eea
As in the Fat Higgs case, this is suppressed by $\Lambda/M_H$.
However, a wide range of parameters
is permitted given the smallness of the observed charged lepton masses.

\subsubsection{Down-type Quarks}

The couplings between the fundamental left-handed quarks
$Q_{1,2}$ to the fundamental right-handed down quarks, $d_{1,2,3}$
is also a dimension five operator.  It can also be generated by the
spectator Higgses, 
\bea
W_{d1} & = & y^d_{ij} \ov{H}^\prime Q_{i} d_j 
\nonumber \\
& \rightarrow &
\left( \frac{y_{H^\prime} y^d_{ij}}{4 \pi} \frac{\Lambda}{M_H}\right)
\ov{H} Q_{1,2} d_i ~.
\label{eq:yd1}
\eea
We also need couplings between $Q_3$ and $d_i$, in order to have a bottom quark
mass.  This requires a dimension 6 interaction between preons, to connect
$Q_3$ and $\ov{H}$ (both mesons) to $d_i$.  This can be arranged by 
integrating out both
$P^\prime$ and $H^\prime$, through the interactions,
\bea
W_{d2} & = & y_{\ov{H}^\prime} \ov{H}^\prime \ov{P}_2 P^\prime
+ y_{d_j} \ov{P}^\prime P_3 d_j
\nonumber \\
& \rightarrow &
\left(  y_{H^\prime} y_{\ov{H}^\prime} y_{d_j} \frac{\Lambda^2}{M_P M_H}
\right) \ov{H} Q_3 d_i ~.
\eea
Note that the NDA estimates do not include a $4\pi$ suppression in this case,
which might point to bottom being naturally heavier than down or strange.
At this point, the down-type quark mass matrix is generic - it contains no
necessarily zero or very small entries.  Thus, it is able to generate all of
the down-type masses, and (after we generate the up and charm quark masses,
below) is sufficient to generate the full CKM structure of the Standard Model.

\subsubsection{Up-type Quarks}

Finally, we need a mass for the up and charm quarks, the top quark having
already been arranged through the dynamical super-potential.  Since the CKM
mixing has already been arranged in the down-type sector, we do not pursue
masses linking $Q_3$ with $u_{1,2}$ (or $Q_{1,2}$ with $t_R$) but instead
just masses connecting $Q_i$ with $u_j$ where $i,j=1,2$.  These can
be generated by integrating out both $P^\prime$ and $H^\prime$,
\bea
W_u & = & y^u_{ij} H^\prime Q_{i} u_{j}
+ y_{P_1} \ov{P}^\prime \ov{P}_1 \ov{P}_{\widetilde{1}}
\nonumber \\ & \rightarrow &
\left( \frac{ y^u_{ij} y_{\ov{H}^\prime} y_{P_1}}{4 \pi} 
\frac{\Lambda^2}{M_P M_H} \right) H Q_{1,2} u_{1,2}
\eea
And thus all Yukawas can be built by integrating out the spectator
preons $P^\prime$ and Higgses $H^\prime$.

\subsubsection{Residual Interactions}
\label{sec:residual}

In addition to the light fermion Yukawa interactions described above, there are
residual effects from integrating out the spectators $H^\prime$ 
and $\ov{H}^\prime$.  The first is that these massive fields mediate
flavor-violating interactions of the form,
\bea
W_{\not F} & = & \left( \frac{y^u_{ij} y^d_{kl}}{M_H} \right) 
Q_i u_j Q_k d_l
+ \left( \frac{y^u_{ij} y^e_{kl}}{M_H} \right) Q_i u_j L_k e_l
~.
\eea
While not a consequence of the composite sector in our model, these types of
interactions are often referred to as ``compositeness operators''
\cite{Eichten:1983hw}.  They lead to interactions involving two SM fermions 
and two of their scalar superpartners, and thus to anomalous flavor 
violation at the loop level.  Given the large value of 
$M_H \gtrsim \Lambda \sim 1000$ TeV, they are not expected to be in 
contradiction with data, though they are in the region where improved 
precision in future experiments could potentially see some of their effects.

The second operator is an induced $\mu$-term for the composite EWSB Higgses
$H$ and $\ov{H}$,
\bea
W_{\mu} & = & \left( y_{P_1} y_{\ov{H}^\prime} y_{H^\prime} 
\frac{\Lambda^3}{M_H M_P} \right) \; H \ov{H} 
\equiv \mu \; H \ov{H} ~.
\eea
As we saw above, a large $\mu$ term would lead to EW fine-tuning, and so we
assume that the Yukawa interactions and/or 
the suppression from $\Lambda / M_H$
is sufficient to bring this operator down to the weak scale.

Both of these features are a consequence of our having taken a minimal
approach to the question of flavor, and not an ``over-kill'' approach
as proposed in \cite{Murayama:2003ag}.  There is no problem to incorporate
the over-kill framework 
in our $SU(3)$ model, though since the contributions are not
sizable enough to be dangerous, we choose to present the simpler and 
potentially more phenomenologically interesting case here.

\subsection{Exotic Quark Masses}

We have already seen that the VEV for the singlet $S$ generates weak scale
masses for the $\psi$ and $\chi$ superfields for fairly
generic parameters.  We also need
masses for the exotic quarks $q_1$, $q_2$, in order to avoid having these
them appear at low energies.  We introduce fundamental fields
$\ov{q}_{(1,2)}$ to marry these exotics through the super-potential,
\bea
W_q & = & y_{q_1} \ov{q}_1 P_3 \ov{P}_1 +
y_{q_2} \ov{q}_2 P_3 \ov{P}_{\widetilde{1}} \nonumber \\
& \rightarrow & \left( \frac{y_{q_1}}{4 \pi} \Lambda \right) \ov{q}_1 q_1
+ \left( \frac{y_{q_2}}{4 \pi}  \Lambda \right) \ov{q}_2 q_2
\label{eq:qmass}
\eea
where we continue to include the NDA $4 \pi$ estimates.  Thus, we typically
expect that $q_1$ and $q_2$ are the heaviest of the exotics.

\section{Unification}
\label{sec:uni}

One of the hallmark successes of the MSSM is 
the prediction of the unification of the gauge couplings.  In this section
we demonstrate that this success can also be preserved in our $SU(3)$ FH
model.  Unlike the generations of the MSSM, our preons do not fill out
complete $SU(5)$ representations, and so it is clear that the standard
structural successes of four dimensional GUTs is not present.  However,
it may be that unification of couplings results from ``string unification''
or from a higher dimensional theory with orbifold GUT breaking 
\cite{Kawamura:1999nj}, in which case matter need not fill out complete 
representations.

The evolution of the gauge couplings takes place in two steps.  Below the
strong coupling scale $\Lambda$ the matter content is that of the MSSM,
including the composite Higgses and top quark,
plus the weak scale exotics $S$, $\psi$, $\ov{\psi}$, $\chi$, and $\ov{\chi}$.
The fields $S$, $\chi$, and $\ov{\chi}$ are singlets under the MSSM gauge
groups, and thus do not contribute to the evolution of couplings at one loop.
Thus, the couplings evolve as,
\bea
\frac{dg_i}{dt} & = & \beta_i \frac{g_i^3}{16 \pi^2}
\eea
with
\bea
\beta_i & = & \left( -3, 1, 39/5 \right)
\eea
for $( SU(3)_C, SU(2)_W, U(1)_Y )$, and we have normalized the hypercharge
coupling in the usual $SU(5)$ way, $\beta_1 = 3/5 \beta_Y$.

Above the scale $\Lambda$ the evolution includes the extra composites
$q_1$ and $q_2$ (and their partners).  More correctly, one  should consider
the evolution in terms of the preons as the relevant degrees of freedom at
large scales, but the two descriptions are equivalent because of 
holomorphicity.  In order to recover unification of couplings, we 
also include two vector-like pairs of spectator ``unifons'' 
which do not participate in the strong dynamics, and are doublets under
$SU(2)_W$ with no hyper-charge.  Thus, above $\Lambda$ we have,
\bea
\beta_i & = & \left( -2, 3, 9 \right) \, ,
\eea
and combining these together with $\Lambda \sim 1000$ TeV, we find unification
of couplings  at the level of $5\%$ at a scale of $3\times 10^{14}$ GeV. 
Such a low scale of unification could be problematic with respect to
proton stability, but since there is no clear GUT structure the 
usual proton decay mediated by $X,Y$ GUT bosons may not be present 
and could be further evaded by imposing some type of baryonic symmetry.

One might worry that the additional strong dynamics will spoil any true
prediction of unification because of the extra strong dynamics
threshold at $\Lambda$.  
In a supersymmetric theory, this is not a problem because the
holomorphicity of the super-potential demands that the low energy couplings
are determined only by the bare masses of the heavy fields 
\cite{Arkani-Hamed:1997ut}.
Thus, our $SU(3)$ FH theory has true unification at a level comparable to the
MSSM.

\section{Phenomenology}  
\label{sec:pheno} 

This model has some distinctive phenomenology, which helps to distinguish it
from other supersymmetric theories.  The MSSM super-partner 
phenomenology depends (as usual) quite
crucially on the mechanism by which SUSY breaking is communicated to the MSSM
fields, and thus is model-dependent.  In order to avoid EW 
fine-tuning, it is important that the scalar partners of top be no more than
a few hundred GeV.  This requirement, combined with a model of SUSY breaking
at high scales will also favor a gluino mass in this region (see
\cite{Kobayashi:2005mg} for models designed to evade this requirement).  
Stop masses of up to about 200 GeV (depending on decay mode and other 
super-partner masses)
can be found in a variety of decay modes at the
Tevatron \cite{Demina:1999ty}, which 
can also typically discover gluinos provided their mass is less than 400 GeV
\cite{Abel:2000vs}.  The LHC is expected to be sensitive to gluino masses up
to about $2$ TeV \cite{Asai:2002xv}.

\subsection{Higgs}

Including the $S$ superfield, our theory has the additional singlet Higgs 
(containing additional neutral scalars and pseudo-scalars) which
mixes through EWSB with the usual MSSM Higgses.  This rich spectrum
corresponds to various cases of the next-to-minimal supersymmetric standard
model, and has been studied in great detail \cite{Miller:2003ay}.
The mixing with the extra scalar state can lead to reduced $Z$-$Z$-$h^0$
and $W$-$W$-$h^0$ couplings, thus weakening the LEP II direct search
limits.  The fermionic component of $S$ will also mix with the MSSM
neutralinos, leading to a modification of the MSSM neutralino properties
\cite{Franke:2001nx}.

The Higgs responsible for EWSB is generally quite a bit heavier than in the
usual MSSM, because of the large value of $\lambda_H$ which contributes
to the Higgs mass.  For large $m_A$, $\tan \beta \sim 1$ 
and $\Lambda \sim 1000$ TeV, the mass is expected to be around 140 GeV, which
is considerably higher than any reasonable value in the MSSM, and high enough
that decays such as $H \rightarrow W W^*$ will begin to dominate.  
More exotic decay modes such as $H \rightarrow A^0 A^0$ may occur, and
can be very challenging for LHC Higgs searches \cite{Dobrescu:2000jt}.
In addition, large values of $\lambda_H$ can lead to the  charged Higgs
being the lightest one, something that never occurs in the MSSM
\cite{Batra:2004vc,Harnik:2003rs}.  

\subsection{Exotics}

\FIGURE[t]{\includegraphics[width=5.0in]{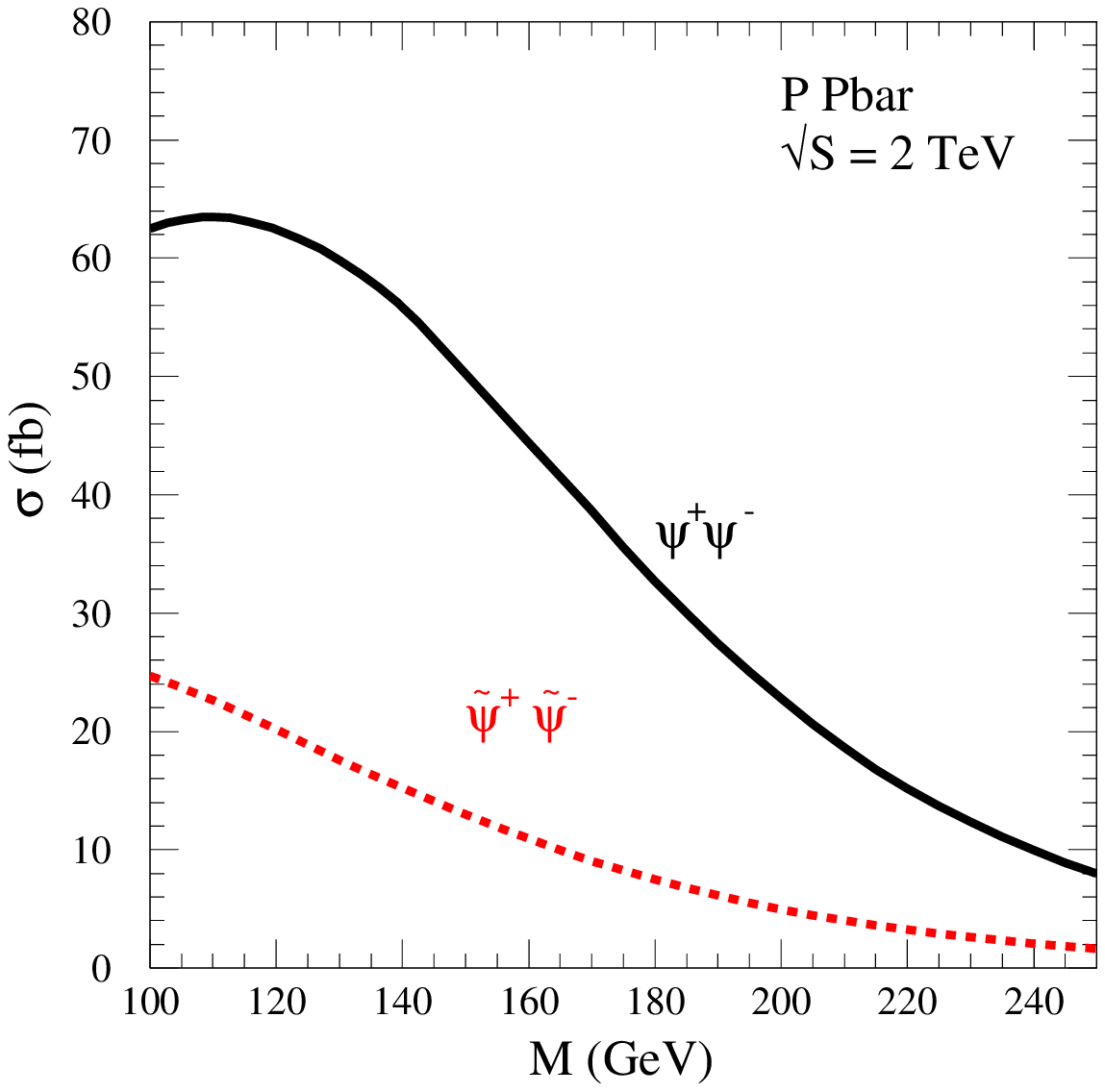}
\label{fig:xsec}\caption{The cross sections for producing $\psi^+ \psi^-$
and $\widetilde{\psi}^* \widetilde{\psi}$ + 
$\widetilde{\ov{\psi}}^* \widetilde{\ov{\psi}}$ at the Tevatron.}}

\FIGURE[t]{\includegraphics[width=5.0in]{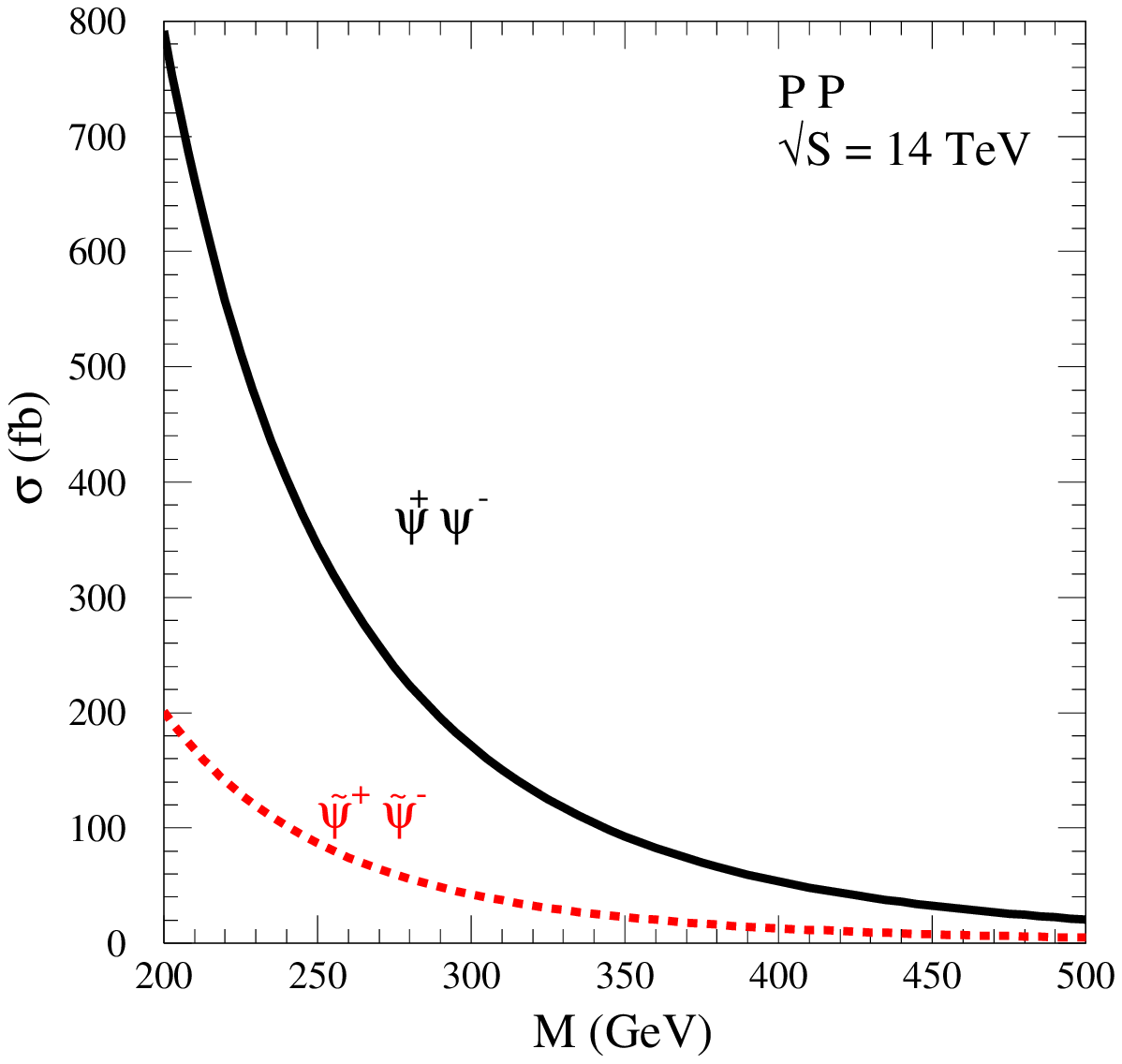}
\label{fig:lhc}\caption{The cross sections for producing $\psi^+ \psi^-$
and $\widetilde{\psi}^* \widetilde{\psi}$ +
$\widetilde{\ov{\psi}}^* \widetilde{\ov{\psi}}$ at the LHC.}}

The model also has a number of additional chiral multiplets.  The colored
quark singlets $q_1$ and $q_2$ have masses of order $\Lambda$
(and thus will probably not be produced at near future colliders), whereas the
color neutral particles are expected to have 
masses $\lambda_\psi \langle S \rangle$, of order $v \sim 200$ GeV.
We expect the lightest of these to be the singlet $\chi$ fields, and the
charge $\pm 1$ fields $\psi$ should be slightly heavier, because of its
non-zero hyper-charge.  We expect that the scalar components will be slightly
heavier than their fermionic partners
because of SUSY-breaking contributions to the scalar masses.

The dynamically generated super-potential has a $Z_2$ symmetry which has all
of the exotic particles coupling in pairs.  This symmetry could be imposed
exactly, but more likely will be broken by interactions such as 
$\ov{q}_1 d_i d_j$, which allows the scalar $\ov{q}_1$ to decay directly 
into down-type quarks (or the fermionic ${q}_1$ to decay into two
quarks and a gaugino).  Since all of the exotic states must decay through
$\ov{q}_1$ whose mass is of order 1000 TeV, the exotics are typically
very long lived and have complicated multi-particle final states.  
In the case of $\psi$, this results in electrically
charged fermions and their scalar partners which are stable on length
scales of the order of the detector, and thus appear as massive
charged objects. Studies in Ref. \cite{Culbertson:2000am} considered
such objects in the context of certain gauge-mediated SUSY breaking models
and conclude that the Tevatron can
discover them with 2 ${\rm fb}^{-1}$
at the $5\sigma$ level provided the production cross section is larger than
about 100 (10) fb for masses of 100 (250) GeV.
In figure~\ref{fig:xsec} we plot the production cross sections for both the
fermion ($\psi$) and scalars ($\widetilde{\psi}$ and $\widetilde{\ov{\psi}}$)
at the Tevatron \cite{Weiglein:2004hn}, through the partonic
processes $q \ov{q} \rightarrow \gamma, Z \rightarrow \psi^+ \psi^-$,
and so forth for the scalars.  Note that the scalar cross sections are
suppressed relative to the fermionic ones because of the intermediate
vector boson, which requires that the scalars be produced in the $p$-wave
to conserve angular momentum.  
For a wide variety of masses, the Tevatron should be able to probe this
scenario with 2 ${\rm fb}^{-1}$ of collected luminosity.
The LHC should be able to produce and
detect the charged quasi-stable particles up to even larger masses.
The cross sections at the LHC are plotted in figure~\ref{fig:lhc}
\cite{Weiglein:2004hn},
and it is expected that the LHC will 
cover the entire parameter space \cite{Ambrosanio:2000zu}.
The $\chi$ and $\ov{\chi}$ particles will be produced much less copiously,
and being electrically neutral and quasi-stable are very difficult to detect.

\section{Conclusions}  
\label{sec:concl} 
  
The Fat Higgs is a fascinating alternative to the minimal supersymmetric
standard model, which may naturally explain why LEP II did not discover the
light CP even Higgs responsible for EWSB.  In this article, we have
examined an alternative to the minimal model based on an $s$-confining
(at $\sim 1000$ TeV)
$SU(3)$ group which generates not only the MSSM Higgses and a singlet,
but also the top quark as composites in the low energy theory.  This
naturally generates the large top Yukawa coupling as a residual of the
strong dynamics, perhaps explaining why top is so much more massive than
any other fermion of the Standard Model.

We are able to generate all of the observed flavor structure of the standard
model, and predict that the Higgs mass and top mass are correlated because of
the common origin of both couplings from the dynamical super-potential.  This
relieves some fine-tuning in the original FH model, and perhaps motivates the
large top mass.  Electroweak symmetry breaking happens in a way which is
reminiscent of the FH, and does impose some mild conditions on the soft masses
of the MSSM-like and exotic Higgses.

The model is compatible with unification of couplings, and
results in some weak scale exotic states not seen the in the MSSM.  These
include quasi-stable electrically charged ($\pm 1$) objects for which there
are good discovery prospects at the Tevatron run II once
2 ${\rm fb}^{-1}$ of data has been collected.  These provide a means to
distinguish this model from other supersymmetric theories, including the
original Fat Higgs itself.  There are also interesting modifications to
Higgs physics, with the most important one being the fact that the lightest
CP even Higgs will typically be heavier than in the MSSM, even at tree level.
Clearly, supersymmetric theories are likely to be richer than even the minimal
models, and the next generation of colliders is likely to have an exciting
time unrevealing the physics at the TeV scale.

\acknowledgments 
 
The authors have benefited from discussions with 
K. Agashe, P. Batra, J. Terning,  H. Murayama, and C.E.M. Wagner, 
and are thankful to R. Harnik, and G. Kribs
for discussions concerning the original Fat Higgs model.  A.D. was partially  
supported by NSF Grants P420D3620414350 and P420D3620434350. 
Fermilab is operated by Universities Research Association Inc.  under  
contract no. DE-AC02-76CH02000 with the DOE.  
Work at ANL is supported in part by the US DOE, Div.\  
of HEP, Contract W-31-109-ENG-38.

\end{document}